# Many Worlds Model resolving the Einstein Podolsky Rosen paradox via a Direct Realism to Modal Realism Transition that preserves Einstein Locality

Sascha Vongehr[†,††]

[†]Department of Philosophy, Nanjing University
[††]National Laboratory of Solid-State Microstructures, Thin-film and Nano-metals Laboratory,
Nanjing University
Hankou Lu 22, Nanjing 210093, P. R. China

The violation of Bell inequalities by quantum physical experiments disproves all relativistic micro causal, classically real models, short Local Realistic Models (LRM). Non-locality, the infamous "spooky interaction at a distance" (A. Einstein), is already sufficiently 'unreal' to motivate modifying the "realistic" in "local realistic". This has led to many worlds and finally many minds interpretations.

We introduce a simple many world model that resolves the Einstein Podolsky Rosen paradox. The model starts out as a classical LRM, thus clarifying that the many worlds concept alone does not imply quantum physics. Some of the desired 'non-locality', e.g. anti-correlation at equal measurement angles, is already present, but Bell's inequality can of course not be violated. A single and natural step turns this LRM into a quantum model predicting the correct probabilities. Intriguingly, the crucial step does obviously not modify locality but instead reality: What before could have still been a direct realism turns into modal realism. This supports the trend away from the focus on non-locality in quantum mechanics towards a mature structural realism that preserves micro causality.

# 1 Introduction: Quantum Physics and Different Realisms

Quantum mechanics has been experimentally confirmed to astounding levels of accuracy. The core of the theory is entanglement. Uncertainty and quantization *can* emerge from classical substrates, but entanglement, which is called superposition if it is the entanglement of states rather than that of multiple particles (which are states of a field), is fundamentally non-classical. All important modern applications like quantum cryptography (Ekert 1991)[1] for example are based on quantum entanglement. Entanglement is proven to be non-classical by the experiments and theory around the Einstein Podolsky Rosen (EPR) paradox (Einstein 1935)[2] and John Bell's famous inequality (Bell 1964)[3].

The violation of Bell inequalities in quantum physical experiments (Aspect 1981; 1982)[4,5] has disproved all local realistic models (LRM), for example non-contextual hidden variables. Such hidden variables cannot violate Bell's inequality (Bell 1966)[6], variations of which (Clauser 1969)[7] have been strongly violated by diverse experiments, most impressively by closing (Weihs 1998)[8] the so called "communication loophole", and quite recently again by confirmation of the Kochen-Specker theorem (Kirchmair 2009)[9]. Desperate attempts at saving localism and unmodified realism try to let LRM exploit the "detection loophole". They have almost retreated to claiming what Abner Shimony calls a *conspiracy*[a] at the intersection of the measurements' past light cones.

---

[a] "… there is little that a determined advocate of local realistic theories can say except that, despite the spacelike separation of the analysis-detection events involving particles 1 and 2, the backward light-cones of these two events overlap, and it is conceivable that some controlling factor in the overlap region is responsible for a conspiracy affecting their outcomes." [Abner Shimony, http://plato.stanford.edu/entries/bell-theorem]

All the above is established beyond doubt, but Bell disproved *local realistic* models. Non-locality is an instantaneous correlation, even though one cannot exploit it to transport matter or information with superluminal velocities. This "spooky interaction at a distance" [b] (A. Einstein) is 'unreal' enough to question the "realistic" in "local realistic" anyway. Zeilinger stresses anti-realism for a number of years, recently with a novel setup (Lapkiewicz 2011)[10]. The Everett relative state description (Everett 1957)[11] is a necessary relativization of terminology that does neither add physics nor necessarily ontological commitment to basic quantum physics. Everett's prose however adopts relative states ontologically, which implies modal realism (Lewis 1986)[12] in philosophy and physicists' many worlds (MW) interpretations (DeWitt 1973; Deutsch 1997)[13,14]. In Deutsch's interpretation, quantum computing is more powerful than classical computing, because the computation is distributed over many 'parallel worlds'[c]. Since they all contribute to the result, all these worlds are real and exist in some concrete sense. Quoting Everett[11] and DeWitt[13], (Kent 1990)[15] has pointed out that MW interpretations (MWI) have realism as their main of two essential characteristics. More recent proponents of MWI are no less insistent on realism. (Tegmark 2007)[16] puts the "External Reality Hypothesis (ERH): There exists an external physical reality completely independent of us humans" prominently onto the first page and even before the mathematical universe hypothesis. Modal realism is alien to the direct realist who believes in one single, classical world that may be deterministic. It is but one step in the ongoing retreat of realism. Many *minds* interpretations (MMI) (Albert 1988; Lockwood 1996)[17,18] will be necessary to tackle the MWI problematic, frequentist probabilities. This 'measure problem' is serious in modern cosmology (Page 2008)[19]. Modifications of realism have been forcefully defended in many places, notably scientifically as well as philosophically stringent by (Saunders 1995; 2008)[20,21] and (Wallace 2005; 2006)[22,23]. However, even just the first necessary step from direct to modal realism has yet to enter the main stream in form of applicable, intuitive models.

---

[b] In 1947, Einstein wrote to Max Born that he could not believe that quantum physics is complete "because it cannot be reconciled with the idea that physics should represent a *reality* in time and space, free from spooky actions at a distance." (Emphasis added) Notice that reality here refers to a directly real world view.

We introduce a simple MW model that resolves the EPR paradox without faster than light ingredients. The model has a surprising and didactically highly valuable twist: Even while already being a MW model, it is still at first a LRM. A single and natural modification turns it into quantum physics. Intriguingly, that local modification obviously modifies *reality* rather than *locality*! This convincing framework is presented as simply as possible in this space and we are confident that a slightly extended version can be understood by advanced high school students and should find its way into undergraduate physics and philosophy lectures.

## 2 The Branching Sausage Many Worlds Model

### 2.1 Many Worlds Models without Entanglement are Classical

Modal realism holds that non-actualized alternatives of random events, like the situation of a fair coin having come up heads while it 'actually' came up tails, are as real as the one actualized relative to the observer who found tails. This is a philosophically self-evident fact; the profound laws of nature do not ensure that I on some Tuesday afternoon find heads instead of tails. Any MW model with 'parallel worlds' constitutes modal realism, but as long as the potential worlds do not interfere, there is no quantum physics and the interpretation of the model as modal realism optional. This will be obvious later when the addition of a single random direction "**DR**" turns our initial MW into a single-world direct realism. It is the entanglement between branches that allows to experimentally show that some of the previously thought 'only potential', like the counter factual in interaction-free bomb-testing (Elitzur 1993; Paraoanu 2006)[24,25], indeed somehow also 'exists'. Classically mutually exclusive states interfere and can exist in superposition inside Schrödinger states (Schrödinger 1935; Lewis 2004; Wineland 2005)[26,27,28]. According to decoherence (Zurek 1998)[29], all states fundamentally stay entangled by the total MW structure; they only decohere for all practical purposes (effectively). Gravity induced 'objective state collapse' (Penrose 1996)[30] insists on strictly not actualized alternatives that will never interfere anymore, while AdS/CFT correspondence argues for unitary quantum gravity. All this is far beyond the scope of the present work and does not seriously impact our MW model either way.

---

[c] 'Parallel' branches describe mostly *orthogonal* states.

**2.2   From Classical 'Meta-Randomness' to Empirical Probability**

Space as such does not reside inside some meta-space. If time is put down as a *t*-axis, it should not be discussed as if there is a 'now-moment' creeping along the axis, as if there is a meta-time that allowed such movement. One rejects such meta-levels, because describing them would require another 'meta-meta' level, leading to infinite regress or at least regress without definite termination. However, when it comes to probability, this error is still common. Consider the branching tree of the potential outcomes of coin tosses. There is no meta level on which we throw a fair meta-coin whenever we reach a branching point. Classical probability equals a 'meta-probability' that can be represented as a phase space volume $V$. A random vector **DR** (for "Direct Realism") may select the actual outcome without bias for any points in that space: The more volume a branch has, the more likely it will be selected. Statistical mechanics similarly assumes fair meta-probability via the ergodic hypothesis. Instead of the whole being already fully described by the tree alone, the meta-probability $V$ makes **DR** behave properly. Such meta-coin tosses are unnecessary and lead to difficulties especially in cosmology, where also space-in-space and time-of-time are most problematic.

In a true MW model, all outcomes are actualized relative to their own branch. You do not advance into the 'heads' instead of the 'tails' branch with meta-probability $V_{heads} = ½$; both futures exist. Most outcome branches of several tosses will observe close to 50% heads. The probability $P$ of an outcome is proportional to the number $N$ of branches with that outcome, which can only in a classical MW model be replaced by $V$, as our model will clarify. Nothing selects any branches or needs to count the parallel branches in order to establish $P$. The branches remember their past, that is enough.

### 2.3  Many Worlds by Cutting a Wiener Sausage

Imagine the EPR experimental setup (See Appendix) embedded inside a straight "Wiener" sausage, i.e. a cylinder with its symmetry axis along the $x$-direction **x**. The sausage volume $V$ is the classical probability. It does not exist in coordinate space, but $V$ splits as dictated by the geometry of that space, thus we depict them in this way. Imagine also a vector **DR** (Fig. 1a) that points to the 'one true real' world which direct realism insists on while taking the rest as mere mathematical construction that reflects the hidden physical mechanisms. When a photon is measured with a beam splitting crystal, the sausage splits according to a right hand rule. The cross product of the arriving photon's propagation direction **p** with the internal *z*-axis of the crystal points toward worlds where the measurement outcome equals zero. At Bob's and Alice's places, this is **x** × **b** and (–**x**) × **a**, respectively. The "dislocations" of decoherence that travel outwards from the measurement events split the world like a "zipper" (Zeh 2010)[31] (Fig. 1a).

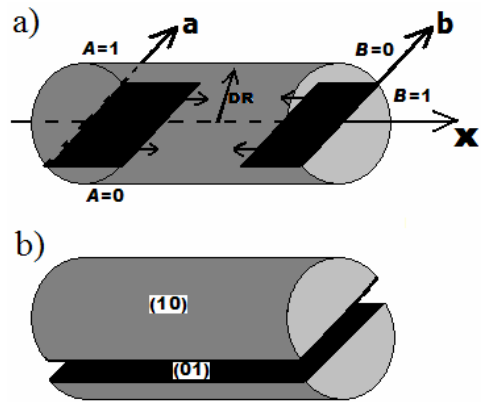

**Fig. 1** World branching in many worlds (MW) models illustrated as literal cutting. a) The cuts are parallel to the measurement directions **a** and **b** and represent the decoherence that propagates with at most the speed of light. Direct realism must assume a direction **DR**; a hidden variable that actualizes one world. b) If the measurement axes are parallel, two parallel worlds result, namely (*AB*) = (10) and (01). In both it seems as if the ends of the sausage knew instantaneously about the respective other end's measurement outcome.

In other words, the measurement vectors **a** and **b** cut the sausage like wires. If parallel, they cut it along the same plane and only two kinds of parallel worlds result (Fig. 1b): One half measured (*AB*) = (01), the other (10).

In order to model arbitrary measurement directions **a** and **b**, let a crystal's internal *y*-direction cut just like its *z*-direction, so that every measurement will split the sausage into four equal pieces. The worlds where the measurement outcome is zero are in opposing quadrants. In Alice's case for example, $A = 0$ worlds are in the first (between her crystal's internal $\mathbf{z} = \mathbf{a}$ and $\mathbf{y}$ directions) and third quadrants (Fig. 2a). The sausage finally falls into eight pieces (Fig. 2b), namely the four different kinds of parallel worlds ($AB$).

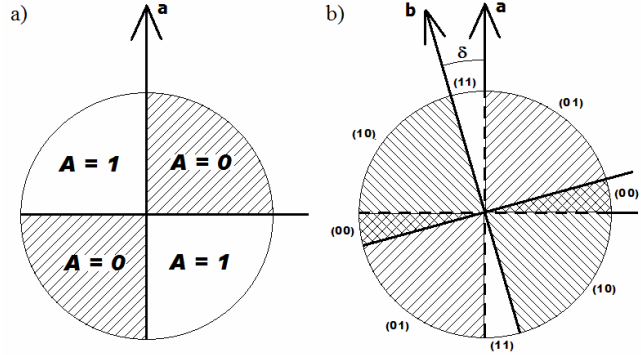

**Fig. 2:** The cross-section of the probability sausage as seen from Bob's perspective: a) Each measurement splits it into four quadrants. b) When both measurements' decoherences overlap, the world has branched into four types of parallel worlds ($AB$). Their volumes $V$ depend on $\delta/\pi$ and cannot possibly mirror the quantum probability $P$ that for example depend on $\sin^2(\delta)$, because this is still fundamentally a local realistic model (LRM) that is proven to not violate Bell's inequality.

If the sausage has unit volume and the angle between **a** and **b** is again $\delta = \beta - \alpha$, the volumes $V(AB)$ of the parallel worlds are $V(\text{E}) = 2|\delta|/\pi$ and $V(\text{U}) = 1 - 2|\delta|/\pi$ (both for $|\delta| \leq \pi/2$ and "E" signifying $A = B$, "U" un-equal). This differs from the quantum expectation values in Eq. (1) [e.g. $N_{ab}(\text{U})/N_{ab} \approx \cos^2(\delta) = 1 - \sin^2(\delta)$] only by the substitution of $2|\delta|/\pi$ for $\sin^2(\delta)$, which coincides at multiples of $\pi/4$. The volumes depend on the relative angle $\delta$ between the cuts. $\delta$ is only known wherever the cuts overlap already, thus the model does harbor the desired 'non-locality' already. However, creating a single hidden variable together with the photon pair, namely the angle $\rho \in [0, 2\pi[$ of the direction **DR** that points to the one real world, proves that this model is a classically deterministic LRM, which cannot violate Bell's inequality. At the Bell angles (See

Appendix Section 3.1), the quantum probabilities *P* violate the inequality maximally: $\cos^2(3\pi/8) + \sin^2(-\pi/4) < \cos^2(\pi/8)$. A truly non-local model could violate the inequality somewhat, but the *V*(*AB*) do not violate it at all: $(1 - 3/4) + (1/2) = (1 - 1/4)$.

Consider now that the sausage is not just split into four branches, but that the MW universe grows new branches instead, which will be crucial. At first, the sausage is empty, having no meat except maybe for tightly along the *x*-axis. Imagine a large number *Z* of new branches, wedge-shaped and labeled by their angle around the *x*-axis, which grow like fiber bundles of meat shooting out of the measurement events. *Z* may be thought of as due to neglected microscopic degrees of freedom. The fibers' angle-specific number density $f = \partial N / \partial \tau$ (with $4 \int_0^{\pi/2} f \, d\tau = Z$ ) depends on the angle from the crystal's *z*-axis, i.e. $f_{(\alpha')}$ at Alice's end. The right hand rule implies that α' increases counter clockwise from the **a** axis; β' increases *clock*wise starting from **b** (Fig. 2b). The new fibers approach each other from both ends just like the cuts did before. Where they meet, the product of the densities integrated over the two *V*(11) parts of the cross-section, that is $2 \int_0^{\delta} f_{(\alpha'=\tau)} f_{(\beta'=\tau-\delta)} d\tau$, may result in a factor proportional to $\sin^2(\delta)$ while the same for the *V*(01) parts is proportional to $\cos^2(\delta)$ with the same proportionality constant. This just needs a suitable choice of the function *f*. β' equals τ – δ, because the integration starts at the **a** direction. $f_{(\beta')}$ does itself *not* know about δ (locality). The shape of *f* is a side issue. For at least three reasons, one cannot yet identify the in such ways resulting factors with the probability: (I) Alice's and Bob's fibers do not match up into continuous (*A*,*B*) parallel world fibers; many are 'left dangling' because $f_{(\tau)}$ is mostly unequal $f_{(\tau-\delta)}$. (II) **DR** still points out the one real world all along, so this is a LRM and no LRM can violate the Bell inequality. (III) When choosing **DR** while creating the photon pair, **a** and **b** are not yet selected. It cannot be rigged to point with a probability that depends on some future $f_{(\rho-\alpha)}f_{(\beta-\rho)}$.

### 2.4 Turning the MW-LRM into a Quantum MWI

Non-locality may suggest modifying the above model by letting Bob's world fiber growth on the right depend on Alice's to the left. Such would bring us back to suspect superluminal hidden information. Instead, we modify the model as naturally expected from the way it was developed up to this point. We forgot that the compound measurement requires another observation, for example Alice's observing of Bob's result, and so the fibers should naturally branch again in order to reflect the fact of that a further observation with several potential outcomes is necessarily involved. The previously considered cutting did so automatically, but with $Z/2$ cutting surfaces on each side approaching, the angle $\delta$ may align some of them into coincident planes like in Fig. 1.

Imagine that on encountering Bob's fibers, the fiber at angle $\rho$ coming from Alice branches into a number proportional to Bob's fibers' density, say $\tilde{Z}f_{(\beta-\rho)}$ new ones. Bob's fibers branch equivalently proportional to the density $f_{(\rho-\alpha)}$ of Alice's fibers. This means that both sides match up exactly into $\tilde{Z}^2 f_{(\rho-\alpha)}f_{(\beta-\rho)}$ new fibers at the angle $\rho$. The shape of $f$ is unimportant. It is not important for us whether the branches first grew according to $f$ and then multiply further according to $f_{(\alpha')}f_{(\beta')}$, or whether they first split into only four worlds on each side which later upon meeting branch into $N(11) \propto (\vec{a} \times \vec{b})^2 = \sin^2(\delta)$, $N(01) \propto (\vec{a} \bullet \vec{b})^2 = \cos^2(\delta)$, and so on. In both and all possibilities in between these extremes of ensuring probabilities that are consistent with the Born rule[d], the introduction of the last, *local* branching accomplishes two crucial aspects simultaneously: (I) It turns the MW-LRM into the correct quantum physical one if only $N(E)/N(U) = \sin^2(\delta)/\cos^2(\delta)$ etc., because the empirical probability will then become for example $P(E) = N(E)/[N(E) + N(U)] = \sin^2(\delta)$. (II) It destroys the direct reality of the model. Before the last branching, **DR** could have pointed all along to a certain future fiber at $\rho$, but it pointed there with probability $V$, not $P$. After the last branching, **DR** does not point to a certain world at all anymore. Just before, it may still point to Alice's fiber $\rho$, but afterwards, it points towards all $\tilde{Z}f_{(\beta-\rho)}$ new ones. A committed direct realist who allows for classical indeterminism may opine that **DR** could randomly select one of the new ones, but since these are all for example (11) fibers, the classical probability $V(11)$ does not change anymore, even if these new fibers are all among each other distinguishable micro-states according to neglected environmental degrees of freedom. The probability 'to go from' a (11)-branch into one of the new (11)-branches is unity, thus $V(11)$ remains what it was before. If Alice at every measurement were to meta-randomly select one from the newly grown fibers, she would end up in $V(11)$ with probability $V(11)$, regardless of how many new branches grow. However, nothing selects. After doing the experiment many times, past experience tells Alice the probabilities, which are in the overwhelming number of worlds the quantum probabilities *P*, not *V*.

---

[d] The second essential characteristic according to Kent is that MWI base the mathematical formalism on a state-vector which belongs to a Hilbert space and has a Hamiltonian evolution. He claims that MWI thus all need a certain axiom that involves continuous time. His main criticism is that the derivation of the Born's rule (probabilities are proportional to the integrated squared amplitudes of the orthogonal wavefunction terms associated with the respective measurement outcomes) must remain a key obstacle for all MWI. We do not presume real continuous parameters and do not *derive* the Born rule. Our claim is that modal realistic local models can *accommodate* Bell violating rules like Born's.

**2.5   Concluding Remarks: Einstein Locality prepared the Modal Paradigm**

If the total number of fibers is to ensure one fiber at the smallest angular resolution, say an angle ε of 0.01 degree, then *V*(11) at $\delta = \pi/2$ would need to grow $1/\sin^2(\varepsilon) > 10^8$ fibers. Without a limit on angular resolution, *Z* is infinite and the cosmological measure problem (Page 2008)[19] has reared its head. We may not be able to normalize the probabilities. Moreover, it is not obvious that the new worlds can be distinguished by neglected microscopic degrees of freedom in such a way that the correct probabilities arise. Such considerations lead us to favor MMI and the view that probabilities are due to what rational agents expect (Deutsch 1999; Wallace 2003)[32,33]. Such is far beyond the scope of this work, which in spite of these known shortcomings of MW models employs them nevertheless in order to address the locality-versus-realism trade-off.

Decoherence "dislocates" (Zeh 2010)[31] via interactions and therefore at most with the speed of light. The model works without superluminal velocities. The last step that turns the model quantum physical is a local branching that destroys the very grounds on which the direction **DR** makes sense. Locality stays; realism is modified. Similar conclusions have been drawn before. The Heisenberg representation of the MWI is local (Deutsch 1999)[34]. However, there are no simple models. The simplicity and the fact that a single, local modification turns the model into quantum physics while destroying its direct realism, is uniquely novel to our approach.

The model clears up a common and tragically consequential confusion that is partially responsible for the slow progress on related issues, for example why it took so long to resolve the EPR paradox. The EPR paradox is traditionally thought of as if quantum mechanics potentially conflicts with special relativity, but this is not just wrong but entirely upside-down. Everett relativity is suspect *without* special relativity. A *non-*relativistic universe would have to quantum split immediately everywhere into extremely many different ones all the time, which seems silly for many reasons. Special relativity already deconstructs the world into a collection of different observers' past light cones in a sort of 'temporal modal realism' – assuming otherwise implies a deterministic block universe. Special relativity is thus prerequisite for understanding quantum mechanics, because world branching only occurs at the observation events while everything outside of one's determined past light cone stays undetermined. Einstein locality and micro causality are important principles in physics – more important even than already widely recognized.

The above conclusion is nicely underlined by our model also teaching that not every MW model is a quantum world and quantum physics is not synonymous with multiverses or modal realism – another in popularity gaining confusion. Without the direction **DR**, which only facilitates didactic, the model is a relativistic MW modal realism all along; only the last step makes it quantum. It is these issues whose surface can only be scratched here that lets us introduce this work as merely one example of a general, thereby highly recommended approach of viewing modal realism as the philosophically self-evident fundament that is already strongly indicated by special relativity, which is best thought about in terms of light-cone descriptions (minds remembering past light cones) rather than hyperspace foliations (slices of a real world). The main relevance of our work lies in accelerating this change of paradigm which we view absolutely necessary for further progress on the foundations of physics.

# 3 Appendix: Pedagogical Introduction to EPR Disproving Local Realistic Models

## 3.1 Basics of the EPR Setup and the Aspect-type Experiment

The simplest version involves a source of pairs of photons. The photons are separated by sending them along the *x*-axis to Alice and Bob, who reside far away to the left and right, respectively. Alice has a calcite crystal polarizing beam splitter with two output channels. Her photon either exits channel "1", which leaves it horizontally polarized, or channel "0", which leads to vertical polarization (relative to the crystal's internal *z*-axis). The measurement is recorded as $A = 1$ or 0, respectively. Bob uses a similar setup, such that there are four possible measurements (*AB*) for every photon pair: (00), (01), (10), or (11). Every pair is prepared in such a way that if the crystals are aligned parallel, only the measurements (01) and (10), short (U) for "Unequal", will ever result. This is called anti-correlation. If the crystals are at an angle $\delta = (\beta - \alpha)$ relative to each other (rotated around the *x*-axis), the outcomes depend on $\delta$ as expected from usual optics at polarization filters: Occurrences of (00) and (11), short (E) for "Equal", increase proportional to $\sin^2(\delta)$. Every experiment starts with the preparation of a pair of photons. When the photons are maybe about half way on their path to the crystals, Alice randomly rotates her crystal either to let $\alpha$ equal $\varphi_0 = 0°$ or $\varphi_1 = 3\pi/8 = 67.5°$. Bob adjusts his crystal similarly to $\beta$ being either $\varphi_1$ or $\varphi_2 = \pi/8 = 22.5°$.

A few didactical points: No other but these "Bell angles" need to be considered for the Bell proof. The magnitudes of δ are multiples of π/8, but one should label with the absolute angles inside any LRM. We consciously avoid probabilities and consider instead expected counts $N$.

Alice and Bob have each only two different angles to choose from, so there are four equally likely combined choices: Out of $N_{\text{Total}} = 160$ runs, the angles are about $N_{ab} \approx 40$ times in each of the four configurations ($\varphi_a$, $\varphi_b$) with $a \in \{0,1\}$ and $b \in \{1,2\}$. The outcomes of all runs are counted by the 16 numbers $N_{ab}(AB)$. The anti-correlation leads to $N_{11}(E) = 0$, while $N_{11}(U) \approx 40$. Generally, it holds

$$N_{ab}(E) \approx N_{ab} \sin^2(\delta), \qquad N_{ab}(U) \approx N_{ab} \cos^2(\delta). \tag{1}$$

Three numbers are important: The sum of $N_{01}(U) \approx 40 * \cos^2(3\pi/8) \approx 6$ and $N_{12}(E) \approx 40 * \sin^2(-\pi/4) \approx 20$ is expected to be by 8 occurrences *less* than $N_{02}(U) \approx 34$ alone.

### 3.2 LRM with Hidden Variables

Let us try to model the experiment described with help of hidden variables. A pair of balls is prepared, say instructions are written on them, and then split. Before the balls arrive, Alice and Bob randomly select angles. Each ball results in a measurement 0 or 1 according to the angle it encounters and the hidden variables it carries. The hidden variables must ensure the anti-correlation at equal angles ($a = b = 1$). Local realism means here that each ball is a real object having all necessary information locally with it. No measurement depends on angles selected far away. This models the fact that photons travel at the speed of light. Nothing travels faster than light, so the photons must know any hidden variables already when they are created and they must bring this information with them on their way.

Assume the instructions somehow prescribe "If $a = 1$, then $A = 0$", or short "$A_1 = 0$". The ball at Bob's place cannot know which angle Alice has just chosen. She *might* have picked $a = 1$, and if so, Bob's measurement cannot be 0 if he also picks $b = 1$. Thus, the hidden variables must prescribe the complementary information "$B_1 = 1$." Furthermore, $A_0$ and $B_2$ must be somehow prescribed by the hidden variables, otherwise the occurrences $N_{ab}(AB)$ cannot reproduce the $\sin(\delta)$ dependence. In summary, the hidden variables may be an infinite table or a complex formula, but they must at least effectively contain the prescription $(A_0, B_1, B_2)$ with $A_1 = 1 - B_1$. With these three degrees of freedom, each pair of balls falls into one and only one of $2^3 = 8$ different classes, which one may index by $i = 4(1-A_0) + 2B_1 + B_2$, with the total number of pairs being $\sum_{i=0}^{7} N^i = 160$ again. Note that index $i$ is not a power. For example, there will be $N^6$ occurrences of (0, 1, 0).

Every pair encounters one of the four possible configurations of angles, hence for example $N^5 = N^5_{01} + N^5_{02} + N^5_{11} + N^5_{12}$. All choices of angles occur about equally often and the hidden variables cannot bias the choice, because they have not arrived yet when the angles are being decided. Hence, all $N^i_{ab}$ are expected to be roughly equal to $N^i/4$, which seems trivial but is the crucial step where Einstein locality is applied:

$$N^i_{ab} \approx N^i/4 \qquad (2)$$

All the cases counted by $N^0_{ab}$, $N^1_{01}$, $N^1_{11}$, $N^2_{02}$, $N^4_{11}$, $N^4_{12}$, and $N^5_{11}$ lead to measurement $(AB) = (10)$. Equivalently, $N^2_{12}$, $N^4_{01}$, $N^4_{02}$, $N^5_{01}$, $N^6_{02}$, and $N^6_{12}$ correspond to (00), while $N^1_{02}$, $N^1_{12}$, $N^2_{01}$, $N^3_{01}$, $N^3_{02}$, and $N^5_{12}$ to (11). Finally, $N^2_{11}$, $N^3_{11}$, $N^3_{12}$, $N^5_{02}$, $N^6_{01}$, $N^6_{11}$, and the four $N^7{}_{ab}$ correspond to (01). This enumerates all the 32 possible $N^i{}_{ab}$ exhaustively. Reproducing the counters $N_{ab}(AB)$ of Section 3.1 leads to for example $N_{11}(E) = 0$ and $N_{11}(01) = N^2{}_{11} + N^3{}_{11} + N^6{}_{11} + N^7{}_{11}$. Important are the following three: $N_{01}(U) = N^0{}_{01} + N^1{}_{01} + N^6{}_{01} + N^7{}_{01} \approx (N^0+N^1+N^6+N^7)/4$, $N_{02}(U) \approx (N^0+N^2+N^5+N^7)/4$, and $N_{12}(E) \approx (N^2+N^6+N^1+N^5)/4$. Bell's inequality is here the mathematically trivial statement that $N^0+N^2+N^5+N^7$ is by $2(N^1+N^6)$ smaller than $N^0+N^1+N^6+N^7$ and $N^2+N^6+N^1+N^5$ all combined. In other words, it is expected that:

$$N_{02}(U) \leq N_{01}(U) + N_{12}(E) \qquad (3)$$

Even if the hidden variables are deliberately chosen in cunning ways[e], the inequality is expected to hold true, because it derives from the randomness of the measurement angles leading to Eq.(2). Therefore, the quantum physical experiment described in Section 3.1, where $N_{02}(U)$ alone is expected to be by 8 occurrences *larger* than the right hand sum, cannot be described by any LRM[f].

---

[e] Simply not preparing any $i = 1$ or 6 pairs sets $N^1$ and $N^6$ equal to zero and ensures that the equal sign in Eq.(3) holds. Resultantly, the Bell inequality can be "violated" every second run. It is crucial to stress that quantum theory predicts the inequality to be violated almost every time.
[f] The CHSH inequality is unnecessary, because from a MWI perspective, the detection loophole will be closed by improving detectors, while the communication loophole is crucial: The photon pair creation event C will 'know' all angles simply by being in the same MW branch as those settings, if only the random setting decision's world-branching had sufficient time to arrive at C.